\documentclass{iopart}

\usepackage{graphicx}

\begin{document}
\jl{1}
\title[Morphology of nodal lines]{The morphology of nodal
lines--random waves versus percolation}
\author{G Foltin\dag\footnote{Present address: Institut f\"ur
Theoretische Physik, Heinrich-Heine Universit\"at D\"usseldorf,
Universit\"atsstrasse 1, 40225 D\"usseldorf, Germany}, S
Gnutzmann\S,  U Smilansky\dag}
\address{\dag Department of Physics of Complex Systems,
Weizmann Institute of Science,
Rehovot 76100, Israel\\
\S Institut f\"ur Theoretische Physik, Freie Universit\"at Berlin, Arnimallee 12,
14195 Berlin, Germany}
\begin{abstract}
In this paper we investigate the properties of nodal structures in
random wave fields, and in particular we scrutinize their recently
proposed   connection with short-range percolation models. We propose
a measure which shows the difference between monochromatic random waves, which are
characterized by long-range correlations,  and Gaussian fields with
short-range correlations, which are naturally assumed to be better
modelled by percolation theory. We also study the relevance of the
quantities which we compute to the probability that nodal lines are in
the vicinity of a given reference line.
\end{abstract}
%03.65.Ge Solutions of wave equations: bound states
%05.40.-a Fluctuation phenomena, random
%05.45.Mt Quantum chaos; semiclassical 

\section{Introduction}
The nodal domains of a (real) wavefunction are regions of equal sign, and are bounded by the nodal lines where the wavefunction vanishes. Even a superficial look at the nodal domains of a quantum wavefunction reveals the  separable or chaotic nature of the quantum system \cite{Str79}.  In separable systems, one observes a grid of intersecting nodal lines, and consequently a checkerboard-like nodal domain pattern. In (quantum) chaotic systems on the other hand, the nodal domains form a highly disordered structure, resembling the geometry found in critical percolation.
Blum \textit{et al} \cite{Blu02} argued that also the statistics of the number of nodal domains reflects the fundamental difference between separable and chaotic quantum systems.  
Bogomolny and Schmit \cite{Bog02} conjectured that the nodal domain statistics of chaotic wavefunctions in two dimensions can be deduced from the theory of critical percolation. They built a percolation model for the nodal domains which allowed them to calculate exactly the distribution of numbers of domains.
Its predictions
have been confirmed numerically as far as nodal counting and the
area distribution of nodal domains are concerned.
While quantum wavefunctions display  long-range correlations,
the critical percolation model assumes that such correlations can
be neglected on distances of the order of a wave length. One may
thus expect that \textit{some} nodal properties in real  wavefunctions are not
well described by critical percolation. The main
motivation of the present study was to investigate the limits of
applicability of the short-range percolation model.
The object we will address is related to the distribution of shapes of nodal lines in the random wave ensemble. To be precise, we will calculate approximately the probability, that a nodal line matches a given reference line up to a given precision $\epsilon$.
%We will compare the predictions for nodal lines  a simpler model for fluctuating lines, which indeed belongs to the percolation universality class. 

\begin{figure}
\begin{center}
\includegraphics[width=0.9\textwidth]{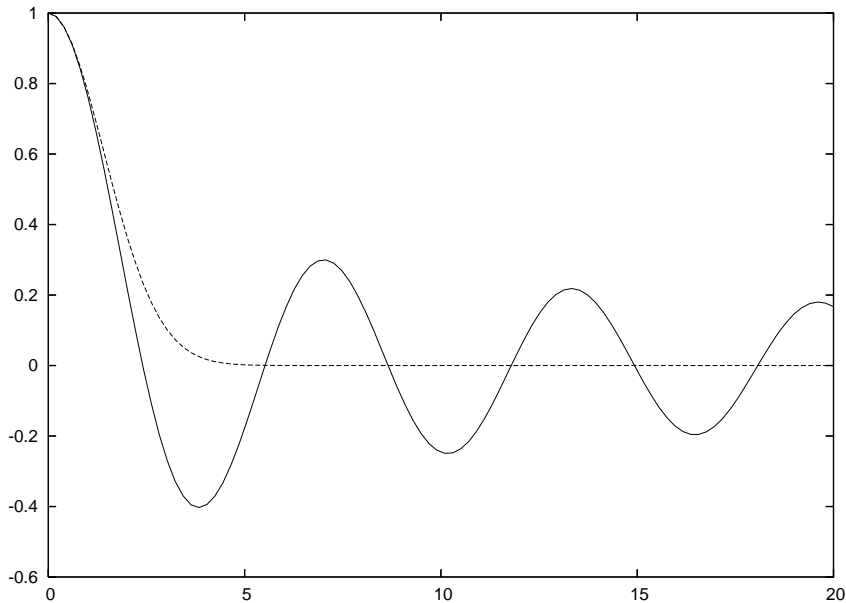}

\caption{The spatial correlation functions $G_1$ (solid), and $G_0$ (dotted) as a function of $kr$.}
\label{corrfu}
\end{center}
\end{figure}

We will examine the statistics of nodal lines within the monochromatic random wave model, which is a good description of the eigenfunctions of a quantum billiard in the semiclassical limit 
\cite{Ber77}.
The monochromatic random wave ensemble consists of solutions of the Helmholtz wave equation for a fixed energy $E=k^2$
\begin{equation}
-\nabla^2\Phi=k^2\Phi.
\end{equation}
Furthermore the random function $\Phi$ is picked up from a Gaussian distribution, i.e. higher order correlations of $\Phi$ can be expressed through the two-point correlation function
$G_1(r)=\left<\Phi(\bi{r})\Phi(\bi{r}')\right>$ by virtue of Wick's theorem.
A convenient representation of $\Phi$ is given by the superposition of cylindrical waves with Gaussian distributed amplitudes
\begin{equation}
\Phi(r,\theta)=\sum_mA_mJ_m(kr)\exp(im\theta)
\end{equation}
where $J_m(x)$ are the Bessel functions of the first kind, and $r,\theta$ is the position in polar coordinates. The Gaussian random variables obey $A^*_m=(-1)^mA_m$ to render $\Phi$ real, and have correlations 
$\left<A^*_mA_{\bar{m}}\right>=\delta_{m,\bar{m}},\,m,\bar{m}\ge0$.
Using the addition theorem for the Bessel functions, one finds for the two-point correlation function
\begin{equation}
G_1(r)=\left<\Phi(\bi{r})\Phi(\bi{r}')\right>=J_0(k|\bi{r}-\bi{r}'|)\sim
\frac{\cos(k|\bi{ r}'-\bi{ r}|-\pi/4)}{\sqrt{k|\bi{ r'}-\bi{ r}|}}.
\end{equation}
It displays in fact long-range correlations, which decay with a power law.
In order to access the relevance of the long-range correlations, we compare the monochromatic random wave ensemble with another Gaussian ensemble of random functions, which, however, does not have long-range correlations, and is characterized by the correlation function
\begin{equation}
G_0(r)=\exp(-k^2r^2/4).
\end{equation}
For the latter ensemble the applicability of the critical (short-range) percolation picture is evident \cite{Zal71,Wei82b}, since the sign of the random function is not significantly correlated for distances $r\gg k^{-1}$.  
Figure \ref{corrfu} shows the spatial correlation functions $G_0, G_1$.

We now briefly introduce the central object of this article. A detailed derivation will be given in section \ref{section:diff}.
Consider a smooth, closed reference curve $\bi{r}(s)$ in the plane, which is parametrized by its arclength $s$.
The integral of the square of the amplitude of a random function $\Phi(\bi{r})$ along this curve
\begin{equation}
X=\frac{1}{2}\int\rmd s\,\Phi(\bi{r}(s))^2
\end{equation}
is itself a random variable. It samples the function not only at a discrete set of points, but along a one-dimensional subset of the plane. It should be well suited to detect the long-range correlations of the random field $\Phi$

Now assume that $\Phi$ has a nodal line very close to the given reference line. Then $X$ will be small in a sense which will be explained later. Thus, by calculating the distribution of $X$, its cumulants or moments, one obtains the relative importance of the given reference line $\bi{r}(s)$.
We will perform these computations for a circular reference line both for the random wave ensemble and for the short-range ensemble defined above. We will study in  particular the scaling properties of the cumulants
of $X$ as functions of the radius (typical size) of the reference curve.
We shall show that they obey a scaling law which distinguishes clearly
between the short-range ensemble and the monochromatic random wave
ensemble. To understand the significance of these cumulants, we shall
consider an \textit{approximate} expression for the probability that a
nodal line is found inside a strip of  width $\epsilon$ about the
reference line. We shall show that this function, when expanded in
powers of $\epsilon^{-1}$ generates the $X$ cumulants. Its scaling
properties with the size parameters, however,  are less sensitive to the
correlations assumed for the underlying random functions model. Strictly
speaking, the function we compute is a measure of the intensity of
fluctuations of the field $\Phi$ along the line, which are certainly
small when a nodal line approximates the reference line, but can also be
small if  different nodal lines which just avoid crossing, are within
$\epsilon$ from the reference line. Since near avoided crossings have
low probability (see \cite{Mon03}) we believe that the function we
compute is closely related to the true probability.

The rest of the paper is organized in the following way. The next
section describes in detail the new concept which we introduce to
the morphological study of nodal lines, that is, the density of
line shapes. Once this is done, a formal expression for the
density, expressed in terms of $X$ is provided, and computed explicitly for particular shapes---circles (section \ref{section:circles}) within a reasonable and calculable approximation. 
These densities are
evaluated for random waves, and for the  short-range ensemble.

\section{The density of nodal line shapes}
\label{section:diff} 
\begin{figure}
\begin{center}
\includegraphics[width=0.7\textwidth]{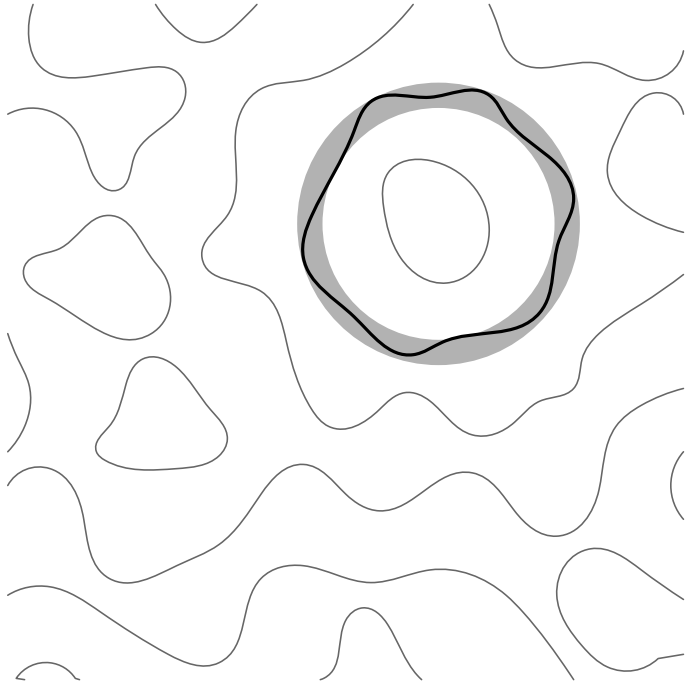}
\caption{A section of the nodal set of a random wavefunction. One of its nodal lines lies within the prescribed thin circular tube, i.e. this configuration contributes to the
density $\rho$.}
 \label{almostcircle}
\end{center}
\end{figure}

We consider two-dimensional, Gaussian random
fields, and a prescribed (closed) reference line. We shall  propose a proper definition of the density of nodal lines
which match the reference line in a random field (or equivalently, the
probability that a nodal line with a prescribed form shows up in a
Guassian random field).
Compared to problems, where the density of
(critical, nodal) points of a Gaussian field is calculated \cite{Ber00,Den01a,Fol03c,Den03a}, we
enter here a new dimension  and consider the density of
one-dimensional strings instead of zero-dimensional, point-like
objects. In order to obtain a well-defined and finite theory, we
have to regularize the theory by dilating the reference curve to a
thin tube with constant thickness $d$ and compute the
probability, that a nodal line is completely inside this tube---see Figure (\ref {almostcircle}). Assume now, that a function $\Phi(\bi{r})$ has a nodal line close to a reference curve $\bi{r}(s)$, where $s$ denotes the arclength. The normal distance $\eta(s)$ of the nodal line from the reference curve can be obtained via linearization 
\begin{equation}
\Phi(\bi{r}+\eta\bi{n})\approx
\eta\partial_n\Phi(\bi{r})+\Phi(\bi{r})=0
\end{equation}
yielding
\begin{equation}
\label{firstorder}
\eta=-\frac{\Phi(\bi{r})}{\partial_n\Phi(\bi{r})}.
\end{equation}
The unit vector $\bi{n}(s)$ is normal to the curve $\bi{r}(s)$.  $\partial_n$ denotes the corresponding normal derivative.
The probability, that a nodal line lies in a \textit{sharp} tube $|\eta(s)|<d$ is, although well defined, not accessible by analytical means. At this point we must resort to further approximations, which will
eventually lead us to a tractable model, at the cost of losing the
rigour of the original object defined above. As a first step we
replace the box shaped cross section by a smooth Gaussian and consider instead the expectation value
\begin{equation}
P_\epsilon=\left<\exp\left( -\frac{1}{2\epsilon}\int\rmd
s\, \eta^2\right)\right>
\end{equation}
where $\int\rmd s$ is the line integral along the reference line, and $\epsilon\sim d^3$.
However, even the computation of this quantity poses unsurmountable difficulties. $\eta=\phi/\partial_n\phi$ is a ratio of two (in general non-independent) Gaussian variables, which is itself non-Gaussian. In order to obtain a tractable expression we approximate the integral  
$w=\int\rmd s\,\eta^2$ by a  mean-field type expression
\begin{equation}
w_m=\int\rmd s\,\frac{\Phi^2}{\left<(\partial_n\Phi)^2\right>}=
\int\rmd s\,\frac{\Phi^2}{\left<(\nabla\Phi)^2\right>/2} 
\end{equation} where the latter step requires isotropy of the distribution of the random field $\Phi$. The final approximation for the shape probability now reads
\begin{equation}
\label{final}\fl
P_\epsilon=\left<\exp\left( -\frac{1}{\epsilon}\int\rmd
s\frac{\Phi^2}{\left<(\nabla\Phi)^2\right>}\right)\right>=\det\left(1+\frac{\hat{B}}{\epsilon \left<(\nabla\Phi)^2)\right>/2}\right)^{-1/2}
\end{equation}
or
\begin{equation}
\label{finall}
F(\epsilon)\equiv\log P_\epsilon=-\frac{1}{2}\sum_\mu\log\left(1+\frac{\beta_\mu}{\epsilon\left<(\nabla\Phi)^2)\right>/2}\right)
\end{equation}
where $\hat{B}$ is an integral operator with (symmetric) kernel 
\begin{equation}
B(s,s')=\left<\Phi(\bi{r}(s))\Phi(\bi{r}(s')\right>=G(|\bi{r}(s)-\bi{r}(s')|)
\end{equation}
and $\beta_\mu$ are the corresponding eigenvalues.
The operator $\hat{B}$ is the correlation function of the field $\Phi$, restricted to the given curve. It is positive semi-definite and has a \textit{finite} trace $\int \rmd s B(s,s)=L$, thus its eigenvalues $\beta_\mu\ge 0$ have an accumulation point at zero.
The final expression for the logarithm of probability (\ref{finall}) is the starting point of our investigation. It should reflect  the relevant features of the inaccessible hard-tube probability, and is an interesting object in its own right\footnote{Private communication with J Hannay.}. It takes into consideration the random field $\Phi$ along the whole reference curve $\bi{r}(s)$. We remark here again that the final approximation for the probability only tests whether $\Phi$ is small along the given curve--it is not able to resolve nearly avoided intersections which are placed next to the reference curve.

$F(\epsilon)$ is the generating function for the cumulants of the random variable $X=(1/2)\int\rmd s\,\Phi^2$
\begin{equation}
F(\epsilon)=\log\left<\exp\left(-\tilde{\epsilon}^{\,-1}X\right)\right>=\sum_{\nu=1,2,3\ldots}\frac{1}{\nu!}(-1)^{\nu}\tilde{\epsilon}^{\,-\nu}\left<X^\nu\right>_c
\end{equation}
where the expansion parameter is $\tilde{\epsilon}=\epsilon\left<(\nabla\Phi)^2\right>/2$. It is \textit{also} the generating function of the traces of powers of the operator $\hat{B}$. In fact, expanding $F(\epsilon)$ in terms of $\tilde{\epsilon}$, i.e. for large $\tilde{\epsilon}^{-1}$, one finds
\begin{equation}\fl
F(\epsilon)=-\frac{1}{2}\sum_m\log\left(1+\tilde{\epsilon}^{\,-1}\beta_m\right)=\sum_{\nu=1,2,3\ldots}\frac{1}{2\nu}(-1)^\nu\tilde{\epsilon}^{\,-\nu}\sum_m(\beta_m)^\nu
\end{equation}
Comparing the two expansions, we see that
\begin{equation}
\left<X^{\nu}\right>_c = \frac{\nu!}{2\nu}
\sum_m(\beta_m)^{\nu}.
\end{equation}

As mentioned in the introduction our goal is to compare two different Gaussian random fields in two dimensions with correlation functions $\left<\Phi(\bi{r})\Phi(0)\right>=G(r)=\int\rmd^2p\,\tilde{G}(p)\exp(i\bi{p}\cdot\bi{r})$, namely
\begin{eqnarray}
G_1(r)&=&J_0(kr)\nonumber\\ 
G_0(r)&=&\exp(-k^2r^2/4).
\end{eqnarray}
$G_1$ is the correlation function of the monochromatic random wave ensemble with a sharply defined energy $k^2$. Consequently, it displays long-range correlations. $G_0$ is a typical short-range ensemble.
Note that the $\tilde{G}$ are normalized such that
\begin{eqnarray}
G(0)=\left<\Phi^2\right>=\int\rmd^2p\,\tilde{G}(p)=1\nonumber\\
-\nabla^2G(0)=\left<(\nabla\Phi)^2\right>=\int\rmd^2p\,p^2\tilde{G}(p)=k^2.
\end{eqnarray} This implies an equal nodal line density $\left<|\nabla\Phi|\delta(\Phi)\right>$ for the long, and short-range ensemble.

\section{The density of circular nodal lines}
\label{section:circles}
\begin{figure}
\begin{center}
\includegraphics[width=0.9\textwidth]{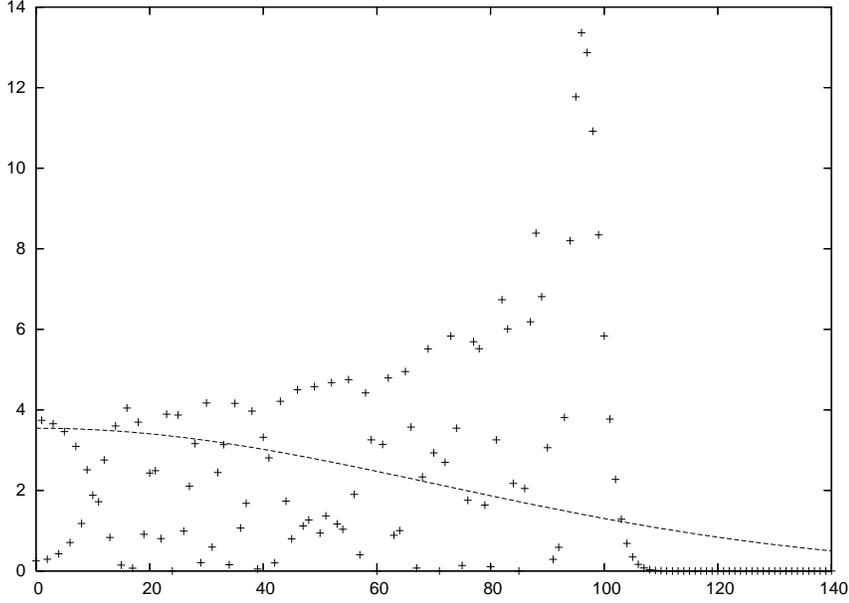}
\caption{The spectrum of $\hat{B}$ as a function of the order $m$ for $kR=100$. Shown is the random wave case (symbol $+$), and the short-range case (points are connected to a dotted line).}
\label{spectrum1}
\end{center}
\end{figure}

We consider now the approximate probability (\ref{final}) for circles with radius $R$.  The kernel of the operator $\hat{B}$ reads for the monochromatic random wave ensemble with correlation function $G_1$ 
\begin{equation}
B(\theta-\theta')=J_0\left(2kR\sin\left(\frac{\theta-\theta'}{2}\right)\right)
\end{equation}
where $\theta,\theta'$ are angles describing positions on the circle.
Owing to the rotational invariance of the problem, the eigenfunctions of $\hat{B}$ are $\exp(im\theta), m=0,\pm1,\pm2,\ldots$.
The eigenvalues of the integral operator are therefore
\begin{eqnarray}
\beta_m&=&R\int_0^{2\pi}\rmd\theta\,J_0\left(2kR\sin\left(\theta/2\right)\right)\exp\left(im\theta\right)\nonumber\\
%&=&2\pi R\sum_{n=-\infty}^\infty\exp\left(-\frac{1-\lambda}{2}k^2R^2\right)I_{m-n}\left((1-\lambda)k^2R^2/2\right)\left(J_n\left(kR\sqrt{\lambda}\right)\right)^2\nonumber\\
&=& 2\pi R\left(J_m\left(kR\right)\right)^2
\end{eqnarray}
The eigenvalues for the short-range ensemble read
\begin{equation}
\beta_m=2\pi R\exp(-k^2R^2/2)\,I_m(k^2R^2/2)\approx\frac{2\sqrt{\pi}}{k}\exp\left(-\frac{m^2}{(kR)^2}\right).
\end{equation}
Figure \ref{spectrum1} shows the eigenvalues of $\hat{B}$ for a circle with radius $kR=100$. The spectrum for the random waves has strong fluctuations, whereas  the spectrum for the short-range ensemble is a smooth (almost) Gaussian.
It was mentioned before, that the trace obeys $\sum_m\beta_m=L$ for both the short- and the long-range ensemble.

\section{The random wave case}
\begin{figure}
\begin{center}
\includegraphics{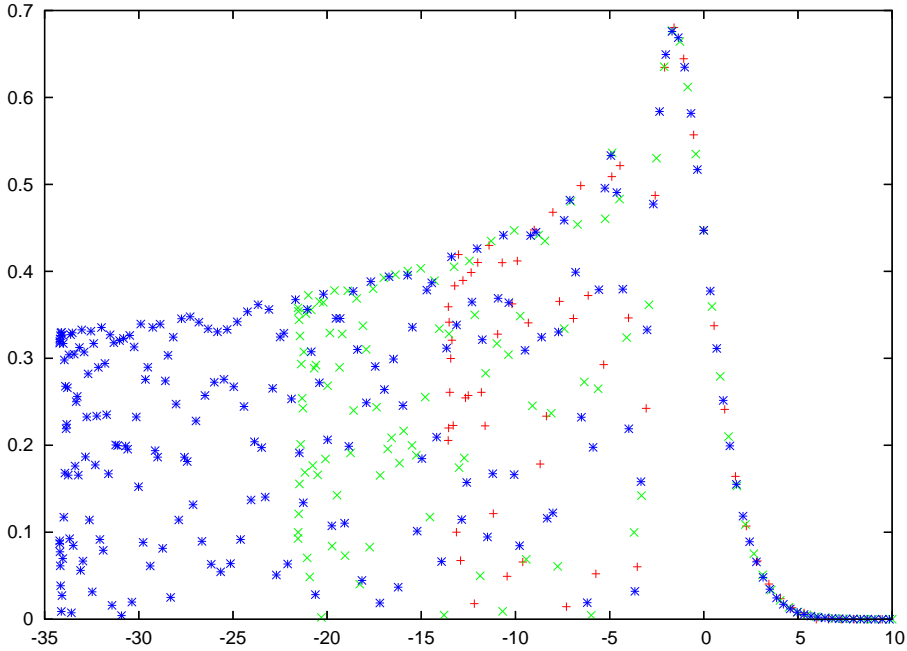}
\caption{The scaling function $f(x)$ for $kR=50$ (red $+$), $kR=100$ (green $\times$), and for $kR=200$ (blue $*$).}
\label{collapse}
\end{center}
\end{figure}

\begin{figure}
\begin{center}
\includegraphics[width=\textwidth]{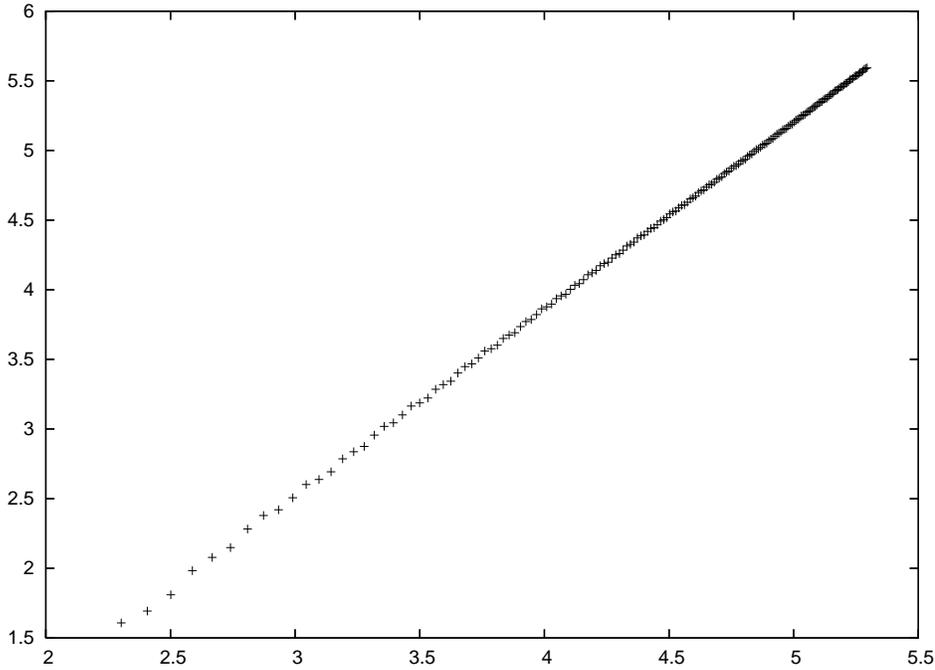}
\caption{Log--log plot of the third cumulant $\left<X^3\right>_c$ as a function of the radius $10<kR<200$ for the random wave case.}
\label{cumulant}
\end{center}
\end{figure}

We calculate now $F(\epsilon)=\log P_\epsilon$ and its  large-$\epsilon$ expansion for large radii $R\gg 1/k$. 
In the region $m<kR$ for large $kR$ the Bessel functions are well approximated by (see \cite{Abr64})
\begin{equation}
J_m(m\sec\beta)\approx\left(\frac{2}{\pi m\tan\beta}\right)^{1/2}\cos\left(m\tan\beta-m\beta-\pi/4\right).
\end{equation}
By setting $m\sec\beta=kR$, we obtain
\begin{eqnarray}
\label{debye}
J_m(kR)&\approx&(2/\pi)^{1/2}\left((kR)^2-m^2\right)^{-1/4}\nonumber\\
&&\times\cos\left(\sqrt{(kR)^2-m^2}-m\arccos(m/(kR))-\pi/4\right).
\end{eqnarray}
In the transition region $m\approx kR$, we approximate the Bessel function $J_m(kR)$ in terms of an Airy function Ai$(x)$ (see \cite{Abr64})
\begin{equation}
J_m(kR)\approx \left(\frac{2}{kR}\right)^{1/3}\textnormal{Ai}\left( \left(\frac{2}{kR}\right)^{1/3}(m-kR)\right).
\end{equation}
We can combine both asymptotic expansions into a scaling law with a universal scaling function $f(x)$
\begin{equation}
\label{scaling}
|J_m(kR)|\sim (kR)^{-1/3}f\left(\frac{m^2-(kR)^2}{(kR)^{4/3}}\right)
\end{equation}
Figure \ref{collapse} shows, that the scaling functions $f(x)$ collapse well for three different values of $kR=50,100,200$. Note that for negative arguments the scaling function $f(x)$ is strongly fluctuating. In the subsequent applications, $f(x)$ and its powers will be integrated over, and for this purpose, $f(x)$ for $x<0$ can be considered as a stochastic function. $f(x)$ vanishes exponentially for $x\to+\infty$, $f(0)=(2/9)^{1/3}/\Gamma(2/3)=0.44731$, and $f(x)\sim (-x)^{-1/4}$ for $x\to-\infty$.
The eigenvalues of the operator scale according to
\begin{equation}
\beta_m=\frac{2\pi}{k}(kR)^{1/3}\left(f\left(\frac{m^2-(kR)^2}{(kR)^{4/3}}\right)\right)^2.
\end{equation}
The leading behaviour of the $\nu$th cumulant of $X$ which is proportional to the trace  of the $\nu$th power of $\hat{B}$,  as a function of the radius $R$ reads
\begin{equation}\fl
\label{moment}
\frac{2\nu}{\nu!}\left<X^\nu\right>_c=\sum_m(\beta_m)^\nu\sim k^{-\nu}(kR)^{1+\nu/3}\int_0^\infty\rmd t\left(f\left((t^2-1)(kR)^{2/3}\right)\right)^{2\nu}
\end{equation}
We find to leading order in $kR$
\begin{equation}
\left<X^\nu\right>_c\sim k^{-\nu}\times\left\{\begin{array}{ll}
kR&\nu<2\\
kR\log(kR)&\nu=2\\
(kR)^{(1+\nu)/3}&\nu>2.
\end{array}
\right.
\end{equation}
This scaling behaviour is compared with the corresponding quantity for the short-range correlations, where
\begin{equation}
\left<X^\nu\right>_c\sim k^{-\nu}\,kR
\end{equation}
for all $\nu>0$. Some remarks are in order.
The cumulants $\left<X^\nu\right>_c$ show a typical critical behaviour for the random wave case. Below the critical power $\nu^*=2$, the large $kR$-scaling does not differ from the short-range case. At the critical power, logarithmic deviations show up, and above $\nu^*$, the cumulant $\left<X^\nu\right>_c$ displays an anomalous scaling in $kR$, different from  the non-critical, short-range ensemble. Now we return to the shape probability
\begin{equation}\fl
\log P_\epsilon\approx-(kR)\int_0^\infty\rmd t\log\left(1+\omega^{-1}(kR)^{1/3}\left(f\left((t^2-1)(kR)^{2/3}\right)\right)^2\right)
\end{equation}
where $\omega=\epsilon k^3/(4\pi)$ is the dimensionless width of the tube around the (here circular) reference curve.
The limit of \textit{small} $\omega$ corresponds to the cumulant (\ref{moment}) for
$\nu\searrow 0$ as far as the scaling behaviour is concerned. Therefore, $\log P_\epsilon$ cannot be considered as a `good' quantity to distinguish between the long-range and the short-range cases---for both ensembles, $\log P_\epsilon\sim kR$. There might be anomalous higher order corrections in the random wave case, which are not considered here. On the other hand, a large-$\epsilon$ expansion (which means arbitrarily wide tubes)
yields the sequence of cumulants (\ref{moment}) for integer $\nu$ which in fact have characteristic scaling properties for $\nu\ge 2$.
Figure \ref{cumulant} shows a log--log plot of the third cumulant ($\nu=3$) as a function of the radius $kR$ for $10<kR<200$. The slope is 1.34063   (standard error $=0.065 \%$), i.e. confirms the predicted exponent $4/3$.

\section{Conclusion}
\label{section:conclusions}
In this paper we compared monochromatic Gaussian random waves and a short-range ensemble of random fields by investigating the statistics of $(1/2)\int\rmd s\,\Phi^2$ along a given reference curve. The $\nu$th order cumulants of this random variable obey  non-trivial scaling laws with respect to the linear size of the reference curve (here circles of radius $R$) in case of the long-range random waves. The second-order cumulant shows logarithmic deviations from the corresponding scaling behaviour of the short-range ensemble. The cumulants of order three and higher have non-trivial exponents. Namely, these cumulants scale like $R^{(1+\nu)/3}$, whereas the cumulants for the short-range ensemble scale $\sim R$.
The probability that a nodal line lies in a circular tube of given thickness $\epsilon$, however, turned out to be a less useful  candidate to probe the long-range properties of the random functions. The logarithm of the shape probability for the short- and the long-range case scale in exactly the same manner, which might explain the success of the ad hoc model \cite{Bog02}.

\ack
We would like to thank N Sondergaard, M Dennis, J Hannay and B Gutkin for many useful discussions.
This work was supported  by the Einstein Center and the Minerva Center for non-linear Physics at the Weizmann Institute, the Israel Science Foundation, and the EU Research Training Network `Mathematical Aspects of Quantum Chaos'.

\appendix

\end{document}